\documentclass[twocolumn,showpacs,showkeys,prc,aps,amsmath,amssymb,10pt]{revtex4-1}
\usepackage{graphicx}
\usepackage{color}
\usepackage{multirow}
\usepackage{bm}% bold math
\usepackage{mathtools}
\usepackage[caption=false]{subfig}
\usepackage{url}
\usepackage{breakurl}
\usepackage{physics}
\usepackage{cancel}
\allowdisplaybreaks

\begin{document}

% Title portion
\title{On the MSW neutrino mixing effects in atomic weak interactions and double beta decays}

\author{Mihai Horoi}
\email{mihai.horoi@cmich.edu}
\affiliation{Department of Physics, Central Michigan University, Mount Pleasant, Michigan 48859, USA}
\date{\today}
\begin{abstract}
 Matter effects on the mixing of the neutrinos mass eigenstates, also know as the Mikheyev-Smirnov-Wolfenstein effect, seem to be  well established in describing the propagation of the neutrino from the source to detecting devices. These effects were mostly considered in bulk matter, 
%such as the Sun or Earth, 
but not inside the atoms. Here we consider the effect of the high electron densities existing in the atomic nuclei. We investigate if these effects can affect the known neutrino phenomenology. 
It was reported that the mixing of the neutrino in high density matter, such as inside a supernova, can affect the Majoron decay probabilities. We investigate if the neutrino mixing effects in the high electron density inside the atomic nuclei can change the neutrinoless double beta decay half-life formula. In both cases we found that the standard results stand. The results look simple, but the road to them is complex and it opens the possibility that the neutrino mixing in atomic nuclei may affect other observables, such as the neutrinoless double beta Majoron decays.
\end{abstract}

\maketitle

\newcommand*{\vv}[1]{\vec{\mkern0mu#1}}

\section{Introduction}

The results of the solar and atmospheric neutrino oscillations experiments were recognized by a recent Nobel prize. The Mikheyev-Smirnov-Wolfenstein (MSW) effect is an essential component needed for the interpretation of these neutrino oscillations experiments\cite{16smirnov}.  Therefore, the mixing of the neutrinos mass eigenstates in vacuum and in dense matter seem to be well established in describing the propagation of the neutrinos from the source to detecting devices. These effects were mostly considered in bulk matter, such as the Sun, Earth, or supernovae, but not inside the atoms. A simple estimation of the electron density and neutrino potential inside a medium-Z nucleus, such as $^{136}$Xe, shows that it is about four orders of magnitude larger than that exiting in the Sun`s core. One could then ask if these high electron densities can produce additional mixing of the mass eigenstates, which needs to be considered in the interpretation of neutrino production and detection phenomenology.

There could be many reasons why the large electron densities existing in the atomic nuclei were not considered (or dismissed). These may be related to the general conditions to have an optical potential, and due the strong connection between neutrino mixing and neutrino oscillations used to describe the solar neutrino problem. However, the main condition actually used is that the neutrino wavelength be smaller than the distance over which the potential/density changes significantly \cite{92giunti1557,03garcia345}. As it will be shown below, this condition is satisfied for a wide range of neutrino momenta and atomic numbers. In addition, Smirnov recently emphasized \cite{16smirnov} the clear distinction between neutrino mixing and neutrino oscillations, and he clearly showed that the high energy solar neutrinos do not exhibit any flavor oscillations, but rather an adiabatic evolution of the mass eigenstates mixing in the solar environment. 
One could also require that the coherence length, related to the size of the wave-packet, be smaller then the  length change in the potential. However, the wave packet itself is not an observable, nor is the process of formation of the wave packet in reactions and decays. Therefore, the coherence length constraint is mostly considered to avoid the details of this process, particularly the situation when part of the wave-packet is in one medium and the rest is in a different medium. Moreover, all cases of interest describing weak decays and reactions are successfully treated using (trains of) plane waves. 
Here, we propose a new approach of treating the evolution of the neutrino mixing due to the neutrino potential generated by the electron density existing inside the atomic nuclei. We consider these effects on individual plane waves contributing to the wave packet, and we show that the standard phenomenology of neutrino emission and detection remains unchanged indicating possible situations were changes may exist and could be tested. 

Neutrinoless double-beta decay ($0\nu\beta\beta$) is considered the best approach to study the yet unknown properties of neutrinos related to their nature, whether they are Dirac or Majorana fermions, which the neutrino oscillation experiments cannot clarify. Should the neutrinoless double-beta transitions occur, then the lepton number conservation is violated by two units, and the black-box theorems~\cite{SchechterValle1982,Nieves1984,Takasugi1984,Hirsch2006} indicate that the light left-handed neutrinos are Majorana fermions. The mass mechanism \cite{84haxton409,Doi1985,87bilenky671,Avignone2008,Vergados2012} is viewed as the mostly likely one contributing to the black box, although other mechanisms were also considered \cite{Vergados2012,11rode1833,Tomoda1991,Doi1985,16ho113,17ho-xv}. Neutrinoless double beta Majoron decay \cite{88doi2575} is an associated process. Majoron decays in supernovae were also analyzed, and it was shown that the decay probabilities are significantly changed \cite{92giunti1557,94berez439} by the additional mixing due to the high matter density. Here, we consider the possible effects to the neutrinoless double beta decay half-life due to the neutrino mixing in the high electron density exiting in the atomic nuclei. We only analyze the mass mechanism, which is the standard one used by the experimental groups to gauge their results. Further analyses of the possible effects to  the $0\nu\beta\beta$ decay probabilities induced by the left-right symmetric model \cite{11rode1833,18ho035502} will be reported separately.

%$0\beta\beta$ reviews: 
%\cite{Vergados2012,11rode1833,Avignone2008,Tomoda1991,87bilenky671,Doi1985,84haxton409}

%Majoron decay:
%\cite{88doi2575,92giunti1557,94berez439}.

%General:
%\cite{99blasone140,92giunti1557,92giunti2414,88mannheim1935,03garcia345,94berez439}.
%See also \cite{giunti2007}.

\section{Neutrino emission and absorption in atomic nuclei} \label{nev}

It is now widely accepted that the flavor neutrinos participating in the weak interaction are coherent superposition of vacuum mass eigenstates. For the neutrino fields, the mixing reads:

\begin{equation} 
\nu_{\alpha L}(x)=\sum_{a=1} U_{\alpha a} \nu_{aL}(x)\ ,
\label{mee}
\end{equation}
where index $\alpha$ indicates a flavor state (electron, muon, tau, $\ldots$), and $a$ designates mass eigenstates (1, 2, 3, $\ldots$). Here the dots indicate sterile flavors, or high mass eigenstates. If one discards the existence of the low mass sterile neutrinos,  the coupling to the higher mass eigenstates is then very small, and the sum over $a$ in  Eq. (\ref{mee}) is reduced to 3. This mixing leads to violations of the flavor number, and it is reflected in the outcome of  the neutrino oscillation experiments. These experiments are mostly  analyzed in terms of neutrino states

\begin{equation} \label{kvm}
\ket{ \nu_{\alpha L}}  = \sum_{a=1} U_{fa}^* \ket{\nu_{aL}}\ ,
\end{equation}
which are dominated by the larger components of the fields. However, some authors treated the neutrino oscillation phenomena by analyzing the effects of the mixing on the fields \cite{99blasone140}, and it was shown that for the ultra-relativistic neutrinos the standard neutrino oscillation results obtained using neutrino states are recovered \cite{99blasone140,92giunti2414}.  In addition, Mannheim \cite{88mannheim1935}, and Giunti and collaborators \cite{92giunti1557} developed a formalism for additional mixing of the field in matter, for either Dirac and Majorana neutrinos. 
Given the justification based on the quantum fields for using the evolution of states, we will first use the states approach to study the evolution of mixing in atomic nuclei. Later in section \ref{dbd}, when analyzing the neutrinoless double beta decay, we will consider the matter mixing of the  neutrino fields in the atomic nuclei.

\subsection{Evolution of mixing in the Sun: two states approximation} \label{sun-ev}

Neutrino states are also used to analyze the matter effects, also known as Mikheyev-Smirnov-Wolfenstein effects \cite{16smirnov}. Here we use the approach of Ref. \cite{03garcia345} (see section III).
For the two-flavor approximation, electron and $X$ (a combination of $\mu$ and $\tau$) in vacuum, one has
\begin{equation} \label{umx}
U \equiv
\begin{pmatrix} 
U_{e1} & U_{e2}\\
U_{X1} & U_{X2}
\end{pmatrix}
=
\begin{pmatrix} 
cos\ \theta & sin\ \theta \\
-sin\ \theta & cos\ \theta
\end{pmatrix}\ .
\end{equation}
In matter, one can use amplitudes \cite{03garcia345} (with probabilities $P_{\nu_e} = |\nu_e|^2$, etc)
\begin{equation} \label{amx}
\begin{pmatrix} 
\nu_{e} \\
\nu_{X}
\end{pmatrix} =
\begin{pmatrix} 
cos\ \theta_m & sin\ \theta_m \\
-sin\ \theta_m & cos\ \theta_m
\end{pmatrix}
\begin{pmatrix} 
\nu_{1}^m \\
\nu_{2}^m
\end{pmatrix}\ ,
\end{equation}
where $\theta_m$ is the total angle that is mixing mass eigenstates to flavor states. One gets
\begin{equation}
cos\ 2\theta_m = \frac{\Delta m^2 cos\ 2\theta-2P V_e}{\sqrt{(\Delta m^2 cos\ 2\theta-2P V_e)^2+(\Delta m^2 sin\ 2\theta)^2}},
\end{equation}
where $P=|\vec{p}|$ is the magnitude of momentum ($P \approx E$ for relativistic neutrinos),  $V_e$ is the coherent potential generated by the charge current interaction between neutrinos and electrons (see below), and $\Delta m^2 = m_2^2 - m_1^2$. For neutrinos in high density electron environments ($V_e > 0$) one gets $\theta_m = \pi/2$, and therefore the electron neutrinos are ``born`` in state 2, while for antineutrinos ($V_e < 0$) one gets $\theta_m = 0$, and therefore the electron antineutrinos are ``born`` in state 1. Generalization to 3 flavors (3 mass eigenstates) indicates that in high electron density media the neutrino (antineutrinos) are ``born`` in the highest (lowest) mass eigenstate. 

For the two state approximation  one gets the following evolution equations for the amplitudes \cite{03garcia345}

\begin{equation} \label{aevolve}
i
\begin{pmatrix} 
\dot{\nu}_{1}^m \\
\dot{\nu}_{2}^m
\end{pmatrix} =
\begin{pmatrix} 
- \Delta(t) & -4i E \dot{\theta}_m(t) \\
4i E \dot{\theta}_m(t)  &  \Delta(t)
\end{pmatrix}
\begin{pmatrix} 
\nu_{1}^m \\
\nu_{2}^m
\end{pmatrix}\ ,
\end{equation}

\noindent 
where the dots denote time derivatives, and
\begin{eqnarray} \label{deltaem}
2P \Delta E_m & \equiv & \Delta(t)  = \mu_2^2(t)-\mu_1^2(t) \nonumber \\
 & = & \sqrt{(\Delta m^2 cos\ 2\theta-2P V_e)^2+(\Delta m^2 sin\ 2\theta)^2}. \nonumber \\
& &
\end{eqnarray}
In eq. (\ref{deltaem}), $\Delta E_m \equiv E_{2m}-E_{1m}$ is the difference of the two neutrinos energies in matter, and
\begin{equation}
\dot{\theta}_m=\frac{\Delta m^2 sin\ 2\theta}{ \Delta(t)^2} P \dot{V}_e\ .
\end{equation}
The general relation between the neutrino optical potential (in $eV$) and the electron density $N_e$ (in $cm^{-3}$) is %(see e.g. \cite{92giunti1557})
\begin{equation}
V_e  = \pm \sqrt{2}G_F N_e \approx \pm 7.6\times 10^{-14}  m_p N_e \ ,
\end{equation}
where the (minus)plus sign corresponds to (anti)neutrinos, $G_F$ is Fermis`s constant, and $m_p$ is the proton mass ($1.67 \times 10^{-24}  g$). Above we used Eq. (2.8) of \cite{92giunti1557}, where the equivalent matter density times the electron fraction $Y_e$ was replaced with $m_p N_e$.
The evolution of the neutrino amplitudes in the Sun for neutrino energies around 10 MeV is adiabatic, i.e. the off-diagonal matrix elements in Eq.(\ref{aevolve}) are small and, therefore, the neutrinos leave the Sun in state 2. For much higher energies the off-diagonal matrix elements dominate, the transition is non-adiabatic, and the neutrinos exit the Sun with same mass eigenstate probabilities ($P_i=\left|\nu_i^m\right|^2=\left|\nu_i\right|^2$) as  when they were ``born`` in the vacuum. Therefore, their detection probability on Earth is the same as for low energy neutrinos that were not affected by the MSW effect. The non-adiabatic transition to the vacuum mixing is not obvious, but it is an outcome of the evolution Eq. (\ref{aevolve}) (see e.g. sections III and IV of Ref. \cite{03garcia345} for a more comprehensive discussion of this phenomenon).

\begin{figure}    
    \centering
\includegraphics[width=0.42\textwidth]{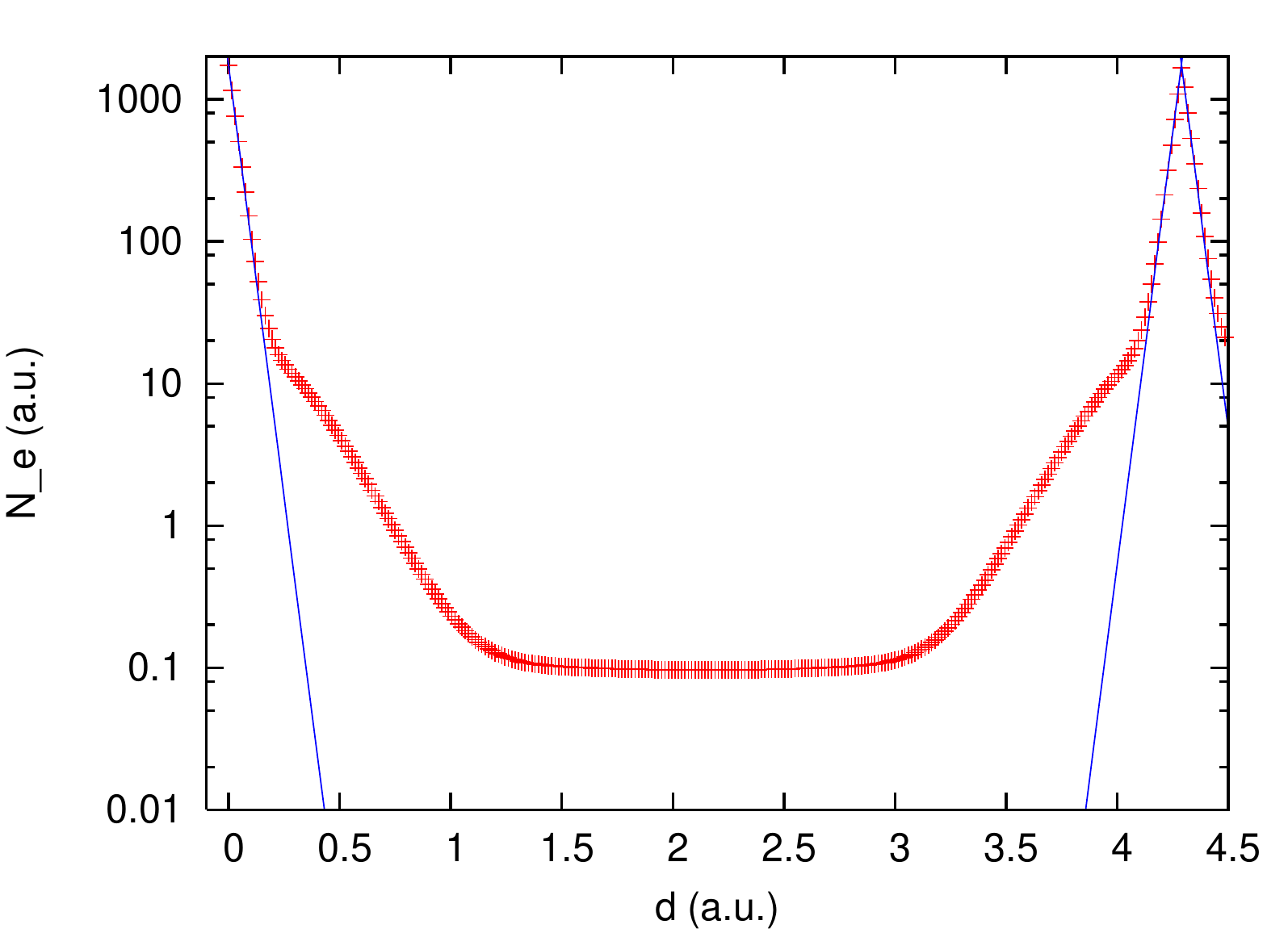} 
\caption{Electron density, $N_e$, in $Si_2$ dimer vs the distance between the two $Si$ nuclei (both in atomic units). The (red) "+" signs are the results of DFT calculations \cite{dftprl1,dftprl2}, and the (blue) continue lines are calculated with Eq. (\ref{rhot}).}
\label{si2}
\end{figure}

\subsection{Evolution of mixing in the atoms} \label{atom-ev}

In atoms, just considering the electron density of two electrons in the lowest s-state of a Hydrogen-like atom (the higher s-states contribute very little $\propto 1/n^3$), one gets
\begin{equation} \label{rhot}
 N_e(t) =  10^{30} \frac{2}{\pi}\left(\frac{Z}{53}\right)^3 e^{-2t Z/53}\ (cm^{-3}),
\end{equation}

\noindent
where $Z$ is the atomic number, and $t$ is in $pm$ ($10^{-12}\ m$). Electron DFT calculations, Fig. \ref{si2}, show that this approximations is very good at and near the nuclei, where the main transition take place (see below). These (equivalent) matter densities at the nucleus for all atoms with atomic number greater than 5 are much larger than those in the Sun`s core. The general requirement for the validity of neutrinos getting mixed and evolving in accordance with the above evolution equation in the optical potential created by a varying electron density is that the neutrino wavelength be smaller than the length over which there is a significant change of the potential \cite{03garcia345,92giunti1557}. In the case of the potential density created by the atomic electron density, this condition reads
\begin{equation}
2\pi\frac{\hbar c}{Pc} \ll \frac{53000}{2Z}\ \ (in\ fm),
\end{equation}
which is satisfied for neutrino energies larger than 2-5 MeV and for a wide range of atomic numbers. This condition is satisfied for all the relevant double beta decay cases, for which the relevant $P$ is of the order of 150 MeV. 
Eq. (\ref{rhot}) indicates that the electron density inside the atomic nucleus is much larger than that in the Sun`s core and, therefore, the (anti)neutrinos are ``born`` in the (lower)higher mass eigenstates.

\begin{figure}    
    \centering
\includegraphics[width=0.42\textwidth]{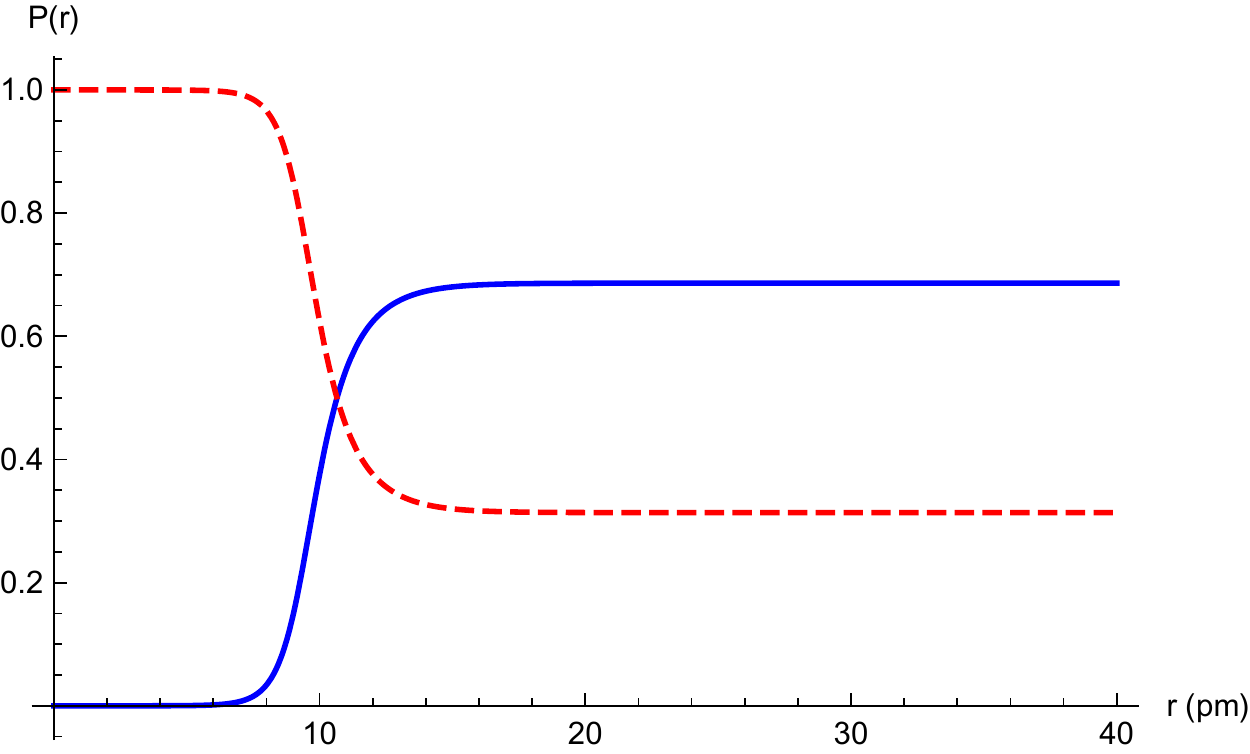} 
\caption{The outgoing evolution of the probabilities of neutrinos produced inside a nucleus (Z=20, P=10 MeV). The neutrinos are produced in state 2 (dashed, red) and they evolve non-adiabatically to 68\% state 1 (full, blue) and 32\% state 2. The horizontal axis represents the distance from the nucleus in $pm$.}
\label{fig1}
\end{figure}

\begin{figure}    
    \centering
\includegraphics[width=0.42\textwidth]{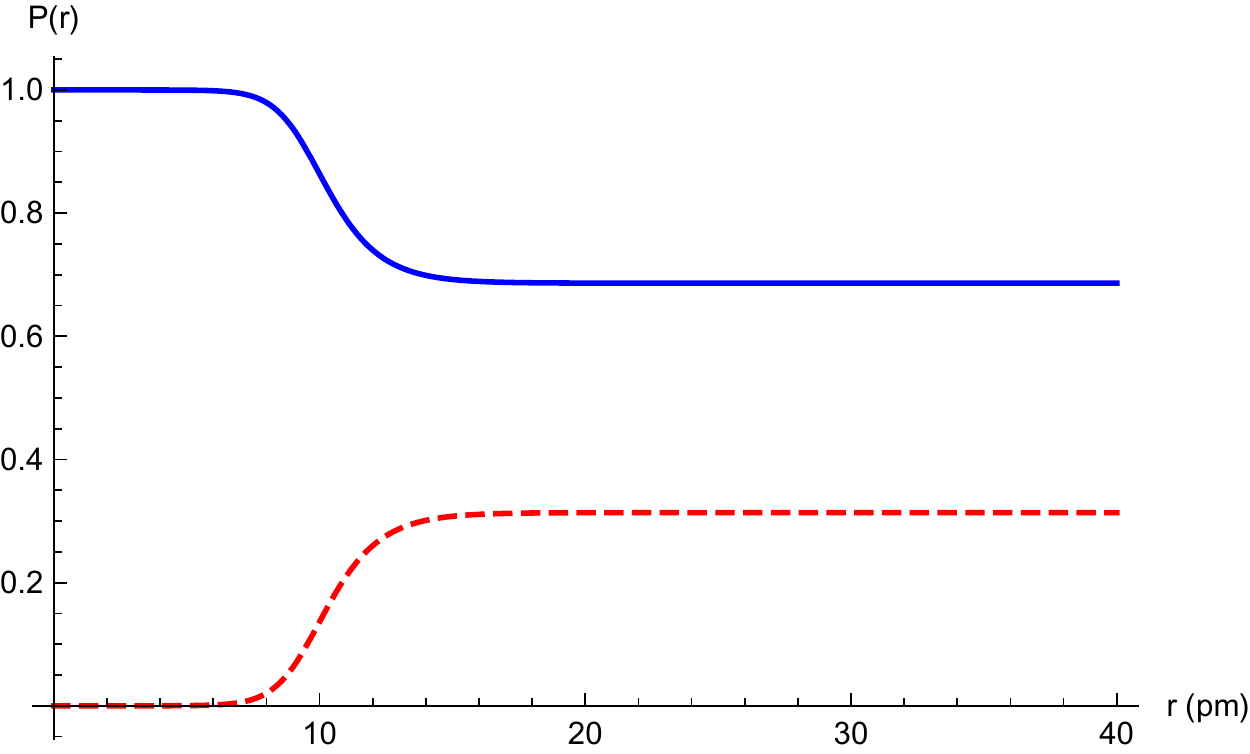} 
\caption{Same as Fig. \ref{fig1} for antineutrinos}
\label{fig2}
\end{figure}

\begin{figure}    
    \centering
\includegraphics[width=0.42\textwidth]{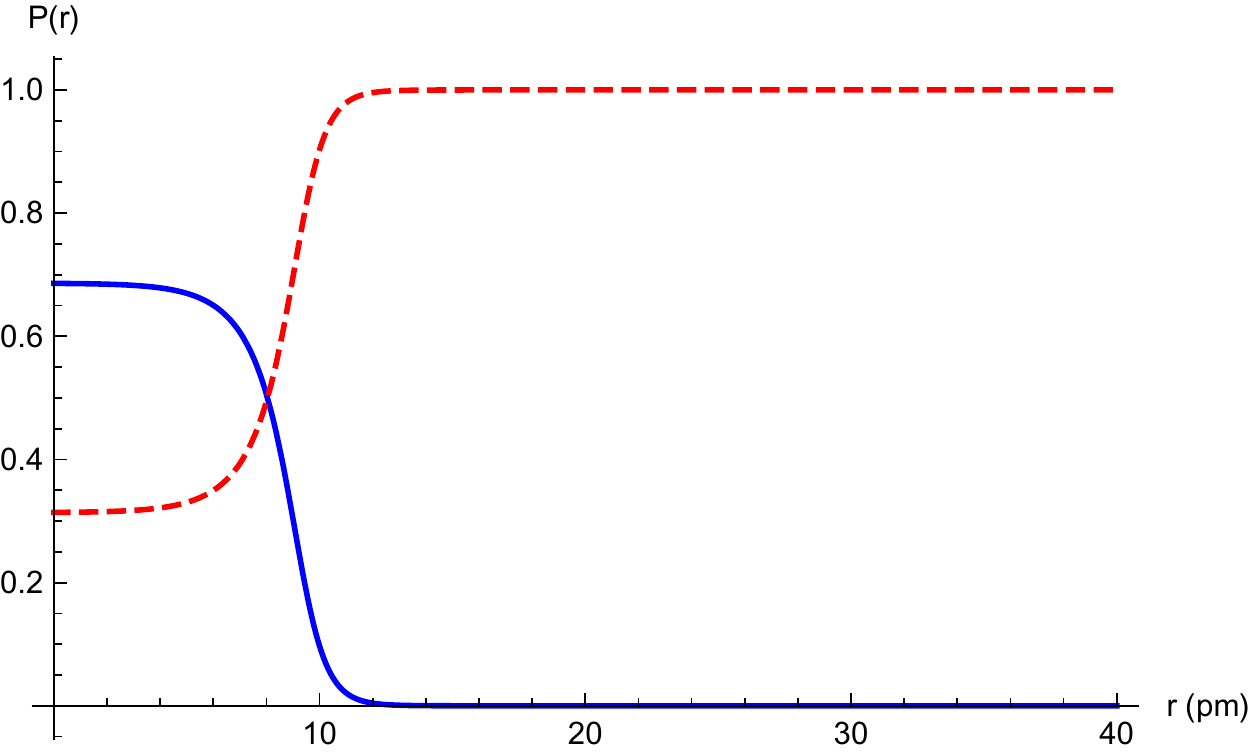} 
\caption{Similar to Fig. \ref{fig1}, but representing a high energy solar neutrino coming in (from right) in state 2 with probability 100\%, which decreases to 32\% when it reaches the nucleus ($r$=0).}
\label{fig3}
\end{figure}

\begin{figure}    
    \centering
\includegraphics[width=0.42\textwidth]{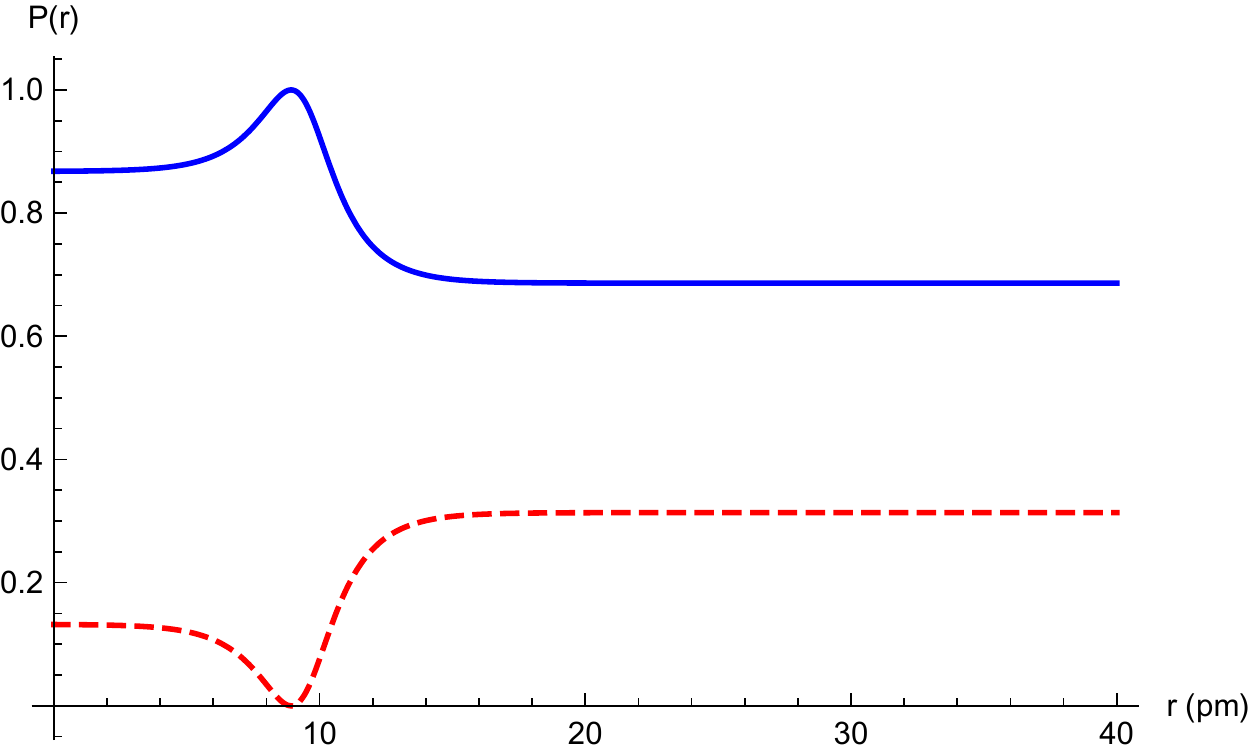} 
\caption{Same as Fig. \ref{fig3} for a regular neutrino (68\% state 1 and 32\% state 2), with phase factor $e^{-i \Delta \phi} = -1$ in Eq. (\ref{ee_ampl}) (see text for details), which arrives with probability 0.13 in state 2 at the nucleus.}
\label{fig4}
\end{figure}

It would be interesting to understand if these matter induced neutrino mixing inside the atomic nuclei can change our understanding of the neutrino detection phenomenology. Let`s first consider the case of an electron neutrino produced with probability 1 as the largest mass eigenstates (2) inside an atomic nucleus (see Fig. \ref{fig1}). Solving the amplitude evolution, Eq. (\ref{aevolve}), one gets for the mass eigenstate probabilities, $P_i \equiv \mid \nu_i^m \mid^2$, the vacuum mass eigenstate probabilities for the electron neutrino ($P_1=cos^2\theta=0.68$ and $P_2=sin^2\theta=0.32$) after passing through the resonance region at around 10 $pm$ from the nucleus. 
Fig. \ref{fig2} presents the similar process for electron antineutrinos ``born`` with probability 1 in state $\ket{1}$ inside the nucleus. The evolution towards the low electron density region (large $r=ct$) brings back the vacuum mass eigenstate probabilities for the electron antineutrino. One can conclude that an antineutrino coming out from the atomic nucleus evolve in the vacuum state that is (up to an overall small complex phase),
\begin{equation} 
\ket{\nu_e} = cos \theta \ket{1} + sin \theta \ket{2}\ ,
\label{exitwf}
\end{equation} 
although it was ``born`` in state $\ket{1}$ inside the atomic nucleus. One should also mention that the neutrino amplitudes in the evolution equation (\ref{aevolve}) are  time dependent multiplicative factors for the neutrino waves. These last ones can be further decompose in spherical components, leading to the usual allowed and forbidden weak interaction rates. 

Considering now the detection process, Fig. \ref{fig3} shows the inward (to the left, nucleus being located in the origin) evolution of a high energy (say 10 MeV) solar neutrino reaching the earth detector that is based on inverse beta decay of a medium mass nucleus (e.g. Cl, Ga, Mo). The neutrino enters the atomic electron cloud in state 2. However, the evolution of the probability for state 2, the only one that can be absorbed inside the nucleus, is reduced to the expected $sin^2\theta$. 

One can also ask how a regular neutrino produced produced in an atomic nucleus on earth (see Eq. (\ref{exitwf})), or a low energy neutrino produced in the Sun will be detected far away from the source. In principle, Eq. (\ref{aevolve}) could be used to further evolve the amplitude on larger scales. However, this is not needed because the evolution in the vacuum starting with state (\ref{exitwf}) is (up to an overall complex phase) well known
\begin{equation} 
\ket{\nu_e} = cos \theta \ket{1} + e^{-i \Delta \phi} sin \theta \ket{2}\ ,
\label{phasewf}
\end{equation}
where $\Delta \phi = \Delta m^2 ct/2P$. The resulting electron neutrino survival amplitude would then be (up to an unimportant overall complex phase)
\begin{equation} \label{ee_ampl}
A_{ee}(t) = \bra{\nu_e(0)} \ket{\nu_e(t)} =cos^2\theta +  e^{-i \Delta \phi} sin^2\theta \ ,
\end{equation}
which leads to the standard neutrino oscillation formula
\begin{equation} \label{ee_prob}
P_{ee}(t) \equiv \left| A_{ee} \right|^2 = 1 - sin^2(2\theta) sin^2(\Delta m^2 L/4P)\ .
\end{equation}

The low energy solar neutrinos (not affected by the solar MSW effect) arrive at the detector with both vacuum amplitudes, but they carry the relative phase factor $e^{-i \Delta \phi}$ (see Eq. (\ref{phasewf})) due to propagation from Sun to Earth. Given that the Sun`s core as a neutrino source is much larger than the oscillation length, the $sin^2(\Delta m^2 L/4P)$ factor in Eq. (\ref{ee_prob}) averages to 1/2 on a large number of neutrino events. Therefore, here we only consider the outcome of Eq. (\ref{ee_prob}) for each event. A general theory of propagation through dense matter and vacuum can be done similar to that in section  III.C of Ref. \cite{03garcia345}. Here we just present some particular cases. For example, Fig. \ref{fig1} viewed in reverse order (right to left) presents the similar evolution for the typical electron neutrino (either produced in the vicinity of the detector or coming from a distant source and arriving at detector with relative phase factor 1 in Eq. (\ref{phasewf}), used as an initial conditions for Eq. (\ref{aevolve})), reaching the detector in the natural mixing of mass eigenstates in the vacuum. One can see again that at the nucleus, the probability of state 2, the only one that is detected inside the high density environment, is 1. This is consistent with the oscillation formula, Eq. (\ref{ee_prob}), when $\Delta \phi$ is a multiple of $2\pi$, the relative phase factor is 1, and $sin^2(\Delta m^2 L/4P)=0$. Fig. \ref{fig4} presents the evolution towards the atomic nucleus of a neutrino arriving at the detector with relative phase factor -1 corresponding to $\Delta \phi$ being an odd multiple of $\pi$. One can easily check using Eqs. (\ref{ee_ampl})-(\ref{ee_prob}) that the survival probability reaches a minimum of $1 - sin^2(2\theta)\approx 0.13$. Fig. \ref{fig4} shows that, indeed, the probability of state 2, the only one that is detected inside the high density environment, is 0.13. We tried several relative phase factors, $e^{-i \Delta \phi}$, in the initial conditions to Eq. (\ref{aevolve}) and we always found results consistent with those of Eqs. (\ref{ee_ampl})-(\ref{ee_prob}).

One can hastily summarize that the two state approximation contains enough elements to conclude that the known neutrino emission and detection phenomenology remains unchanged when the mixing in the atomic nuclei is taken into account. However, the interpretation given in Fig. \ref{fig3} for the detection of the high energy solar neutrino could raise new questions when the three neutrino mixing is considered and the normal mass ordering is assumed. In that case the neutrinos can cross only one resonance in the Sun, arriving at Earth in state 2, while in the high electron density existing in the medium Z atomic nuclei one would identify only state 3 as an electron neutrino. 

\subsection{Beyond the two states approximation} \label{3state}

To resolve this puzzle one needs to consider the MSW evolution for three (or more) mass eigenstates mixing in Eqs. (\ref{umx})-(\ref{amx}). Here we extend the evolution equation (\ref{aevolve}) following the algorithm described in Eqs. (51)-(63) of Ref. \cite{03garcia345}. The vector of 3 flavor amplitudes is denoted as $\nu_f = \left( \nu_e,\ \nu_{\mu},\ \nu_{\tau} \right)^T$, and the vector of matter mass eigenstate amplitudes is denoted as $\nu_m = \left( \nu_1^m,\ \nu_2^m,\ \nu_3^m \right)^T$. Then the Schroedinger-like evolution equation for the flavor amplitudes in matter reads
\begin{equation} \label{wolf}
i\frac{\partial \nu_f}{\partial t} = \left( H_0 + V \right) \nu_f \ ,
\end{equation}
where $H_0 = U^{\dagger} diag \left(m_1^2/(2P),m_3^2/(2P),m_3^2/(2P)\right) U$,  
%$a=1,\ldots,3$,
and $V = diag \left( V_e+V_N, V_N, V_N \right)$. $m_a$ are the vacuum masses of the mass eigenstates, and $V_N$ is the neutral current potential generated by (mostly) neutrons and protons. Using the unitary transformation of the amplitudes in matter 
\begin{equation} \label{amx3}
\nu_f = U_m \nu_m \ ,
\end{equation}
one can write an alternative evolution equation
\begin{equation} \label{uevolve}
\frac{\partial \nu_f}{\partial t} =\dot{U}_m \nu_m + U_m \dot{\nu}_m \ ,
\end{equation}
where the upper dots represent total derivatives with respect to $t$ (time or distance).
Combining Eq. (\ref{wolf}) with Eq. (\ref{uevolve}) one gets
\begin{equation} \label{aevolve3}
i \dot{\nu}_m = U^{\dagger}_m \left(H_0 + V\right) U_m \nu_m -i U^{\dagger}_m \dot{U}_m \nu_m \ .
\end{equation}
Comparing this equation with Eq. (\ref{aevolve}), one concludes that the first term on the right hand side of Eq. (\ref{aevolve3}) corresponds to the diagonal terms in Eq. (\ref{aevolve}), and the second term corresponds to the off-diagonal terms in Eq. (\ref{aevolve}). Following the discussion after Eq. (\ref{aevolve}) one can conclude that in the extreme no-adiabatic conditions existing in medium-Z atoms, the first term in Eq. (\ref{aevolve3}) can be neglected. One can go further and conclude that in the evolution of mixing under the conditions described above one can neglect the term 
$\left(H_0 + V\right) U_m$, which when combined with Eqs. (\ref{amx3}) and inserted in Eq. (\ref{wolf}) one gets
\begin{equation} \label{wolf-na}
i\frac{\Delta \nu_f}{\Delta t} \approx 0 \ .
\end{equation}
The interpretation of this equation is that during the short $\Delta t$ transition, under extreme non-adiabatic conditions existing in medium-Z atoms, the flavor amplitudes do not change, while the mass eigenstate amplitudes may change dramatically. This interpretation needs to be further investigated numerically for lower-Z atoms, where 
%both adiabatic and non-adiabatic evolution of the mixing 
mixed amplitudes may co-exist.

Following the above interpretation on can conclude that whatever is the vacuum electron neutrino amplitude outside the atomic nuclei, it will not change when the neutrino arrives at the nucleus. Similarly, the neutrino created as the electron flavor inside the atomic nucleus will exit the atom as the known mixture of vacuum mass eigenstates described by PMNS matrix.
One can conclude that the effects of mixing in the high electron density existing inside the atomic nuclei, combined with the standard evolution of these mixings through the atomic electron cloud, do not change the known neutrino emission and detection phenomenology.

\section{Neutrinoless double beta decay in atomic nuclei} \label{dbd}

Knowing that the neutrino mixing in matter changes the Majoron decay probabilities \cite{92giunti1557,94berez439}, it would be interesting to find out if the neutrinoless double beta decay mass-mechanism probabilities are affected by the high electron density existing in all relevant double beta decay isotopes. Naively, one could be worried by the simple $0\nu\beta\beta$ interpretation that relies on the fact that one neutron emits a antineutrino (that would be ``born`` in the lowest mass eigenstate inside the atomic nucleus), which will be absorbed as a neutrino (that should be in the highest mass eigenstate).

The part of the neutrinoless double beta decay amplitude relevant to the neutrino fields is the neutrino propagator (NP)

\begin{equation} \label{epropagator}
NP = \bra{0} T\left[ \psi_{eL}(x_1) \psi_{eL}^T(x_2) \right] \ket{0} \ ,
\end{equation}
where $\psi_e(x)$ is a four component Majorana neutrino spinor field. For double beta decay only the left-chiral components of the electron neutrino field contribute. The standard derivation of the $0\beta\beta$ decay half-life assumes that the electron neutrino fields can be expanded in terms of the vacuum mass eigenstates, Eq. (\ref{mee}), and one gets (up to some phases) \cite{87bilenky671}

\begin{eqnarray} \label{4propagator}
NP = & \sum_{a=1}^3 U_{ea}^2 \bra{0} T\left[ \psi_{aL}(x_1) \psi_{aL}^T(x_2) \right] \ket{0} \nonumber \\
= & \sum_{a=1}^3 U_{ea}^2 \left[ -i \int \frac{d^4p}{(2\pi)^4} \frac{m_ae^{-i p \cdot (x_1-x_2)}}{p^2-m^2_a+i \epsilon} P_L {\cal{C}} \right] \ .
\end{eqnarray} 
Here $P_L$ is the left-handedness projector operator, and ${\cal{C}}$ is the spinor charge conjugation operator. The product $P_L {\cal{C}}$ is further used for processing the electron current, and one arrives to the standard formula for the $0\beta\beta$ decay constant \cite{87bilenky671}

\begin{equation} \label{lifetime}
\frac{1}{T_{1/2}} = G(Z,Q) \left| M_{0\nu} \right|^2 \left| \sum_{a=1}^3 U_{ea}^2 m_a \right|^2 / m_e^2 \ ,
\end{equation}
where $G(Z,Q)$ is a phase space factor \cite{16ne748}, $M_{0\nu}$ is a nuclear matrix element \cite{Avignone2008,Vergados2012,13ho014,NeacsuHoroi2016}, and $m_e$ is the electron mass.

If one wants to consider the MSW effects due to the high electron density in the atomic nuclei, one has to take into account that different components of the vacuum mass eigenstate fields in Eq. (\ref{mee}) are changing differently. A simpler approach is to use 2-components spinor fields (see  Refs. \cite{92giunti1557,92giunti2414,88mannheim1935,87bilenky671,giunti2007}). 
Then, one needs to make the connection to the four-components spinor fields necessary to further process the electron current. The 2-component spinor approach to calculate propagators is somewhat more complicated than that for 4-components spinors \cite{10dreiner1,94berez439}. However, in order to avoid the potential error due to different phase convention, we decided to follow on the 2-component spinor formalism developed in Refs. \cite{92giunti1557,giunti2007}. The typical approach is to use a specific representation of the Dirac matrices, the Weyl's chiral representation being the most convenient. Using the phase conventions of Ref. \cite{giunti2007} (see Eqs. (A.109)-(A.122)), in the Weyl's chiral representation one gets 
\begin{equation} \label{plc}
P_L {\cal{C}} = 
\begin{pmatrix}
0 & 0 \\
0 & i\sigma^2
\end{pmatrix}\ ,
\end{equation}
and
\begin{equation} \label{psil}
\psi_L(x) = 
\begin{pmatrix}
0 \\
\Phi(x)
\end{pmatrix}\ .
\end{equation} 
Then, the contributions to the propagator in Eq. (\ref{4propagator}) look like

\begin{eqnarray} \label{42propagator}
& \bra{0} T\big[ \psi_{aL}(x_1)  \psi_{aL}^T(x_2) \big] \ket{0} = \nonumber \\
&
\begin{pmatrix}
0 & 0 \\
0 & \bra{0} T\left[ \Phi_{a}(x_1) \Phi_{a}^T(x_2) \right] \ket{0}
\end{pmatrix}\ .
\end{eqnarray}

By comparing with Eqs. (\ref{4propagator}) and (\ref{plc}) on can conclude that
\begin{equation} \label{2propagator}
 \bra{0} T\left[ \Phi_{a}(x_1) \Phi_{a}^T(x_2) \right] \ket{0} = 
\sigma^2 \int \frac{d^4p}{(2\pi)^4} \frac{m_a e^{-i p \cdot (x_1-x_2)}}{p^2-m^2_a+i \epsilon}.
\end{equation}

Following Ref. \cite{92giunti1557} one can relate the electron neutrino field to the mass eigenstates in matter in two steps. In the first step one can translate Eq.  (\ref{mee}) for four-component spinors to two-component spinors

\begin{equation} \label{phievm}
\Phi^W_{\alpha}=\sum_{a=1}^{N} {\cal{U}}_{\alpha a} \Phi^M_{a} \approx \sum_{a=1}^{3} U_{\alpha a} \Phi^M_{a}\ ,
\end{equation}
and then relate the vacuum two-component spinors to the two-component spinors in matter.
Using the notations of Ref. \cite{92giunti1557}, here and below $\alpha$ represent flavor indexes, $a$ represent vacuum mass eigenstate indexes, and $j$ represent matter mass eigenstate indexes.
This last relation is given in Eq. (2.19) of Ref. \cite{92giunti1557}, and considering the approximation for ultra-relativistic neutrinos of Eq. (2.21) in Ref. \cite{92giunti1557},  one  can write

\begin{widetext}
\begin{eqnarray} \label{phimvm}
 \Phi^M_{a}(x)=\int \frac{d^3p}{(2\pi)^{3/2}} \sum_j & \Big[ \left(\alpha^{(-)}_{a\ j} w(\vv{p},-) a_j(\vv{p},-) + \frac{m_a}{2P} \beta^{(+)}_{a\ j} w(\vv{p},+) a_j(\vv{p},+) \right) e^{-i p \cdot x} \nonumber \\
 & + \left( \frac{m_a}{2P} \alpha^{(-)*}_{a\ j} w(\vv{p},+)a^{\dagger}_j(\vv{p},-) - \beta^{(+)*}_{a\ j} w(\vv{p},-)a^{\dagger}_j(\vv{p},+) \right)e^{i p \cdot x}  \Big]\ ,
\end{eqnarray}
where $\alpha^{(-)}$ and $\beta^{(+)}$ are unitary matrices depending on the magnitude of the momentum $P$ (see Eqs. (2.22)-(2.23) of Ref. \cite{92giunti1557}.

Let`s consider first the case of low matter density, for which the matter effects do not produce any significant mixing. In addition, using the conventions given in Eqs. (A.109)-(A.122) of Ref. \cite{giunti2007}, one can identify the $w(\vv{p},h)$ spinors with the $\chi^{(h)}(\vv{p})$. Then one can write for the field of mass eigenstate $a$

\begin{eqnarray}
 \Phi_{a}(x)=\int \frac{d^3p}{(2\pi)^{3/2}} & \Big[ \left(  \chi^{(-)}(\vv{p}) a_j(\vv{p},-) + \frac{m_a}{2P}  \chi^{(+)}(\vv{p}) a_a(\vv{p},+) \right) e^{-i p \cdot x} \nonumber \\
 & + \left( \frac{m_a}{2P}  \chi^{(+)}(\vv{p})a^{\dagger}_j(\vv{p},-) - \chi^{(-)}(\vv{p})a^{\dagger}_j(\vv{p},+) \right)e^{i p \cdot x}  \Big]\ .
\end{eqnarray}

One can then calculate the contributions to the time ordered product entering  propagator in Eq. (\ref{2propagator}) as

\begin{equation}
\bra{0}  \Phi_{a}(x_1) \Phi_{a}^T(x_2) \ket{0} =
\int \frac{d^3p}{(2\pi)^3}\frac{m_a}{2P} \big[ \chi^{(-)}(\vv{p})  \chi^{(+)T}(\vv{p}) - \chi^{(+)}(\vv{p})  \chi^{(-)T}(\vv{p}) \big] e^{-i p \cdot (x_1-x_2)} \ ,
\end{equation}

\begin{equation}
-\bra{0}  \Phi^T_{a}(x_2) \Phi_{a}(x_1) \ket{0} =
\int \frac{d^3p}{(2\pi)^3}\frac{m_a}{2P} \big[ \chi^{(-)}(\vv{p})  \chi^{(+)T}(\vv{p}) - \chi^{(+)}(\vv{p})  \chi^{(-)T}(\vv{p}) \big] e^{i p \cdot (x_1-x_2)} \ .
\end{equation}
\end{widetext}
Given that the neutrinos are ultra-relativistic one can replace $2P$ in the denominators with the on-shell energy $2E_p$ that enters the standard derivation of Feynman propagators \cite{peskin1995,srednicki2007}. Therefore, to be consistent with Eq. (\ref{2propagator}) one needs to have
\begin{equation}
\chi^{(-)}(\vv{p})  \chi^{(+)T}(\vv{p}) - \chi^{(+)}(\vv{p})  \chi^{(-)T}(\vv{p}) = -i \sigma^2 \ ,
\end{equation}
which can  be also checked directly.

Now we can go back to the electron neutrino fields in high electron density environments. By combining  Eq. (\ref{phievm}) and (\ref{phimvm}) one gets

\begin{widetext}
\vspace{1cm}

\begin{eqnarray} \label{phief}
 \Phi^W_{e}(x)=\int \frac{d^3p}{(2\pi)^{3/2}} \sum_{a, j} U_{ea} & \Big[ \left(\alpha^{(-)}_{a\ j} \chi^{(-)}(\vv{p}) a_j(\vv{p},-) + \frac{m_a}{2P} \beta^{(+)}_{a\ j} \chi^{(+)}(\vv{p}) a_j(\vv{p},+) \right) e^{-i p \cdot x} \nonumber \\
 & + \left( \frac{m_a}{2P} \alpha^{(-)*}_{a\ j} \chi^{(+)}(\vv{p})a^{\dagger}_j(\vv{p},-) - \beta^{(+)*}_{a\ j} \chi^{(-)}(\vv{p})a^{\dagger}_j(\vv{p},+) \right)e^{i p \cdot x}  \Big]\ .
\end{eqnarray}
Some of the contributions entering the neutrino field can be simplified if one considers the high electron density medium where the neutrinos are ``born``. In that case the terms without masses reduce to one state (see e.g. Eq. (3.21) of Ref. \cite{92giunti1557} , and Eqs. (2.25) and (2.34) of Ref. \cite{88mannheim1935})
\begin{eqnarray}
\sum_a U_{ea} \alpha^{(-)}_{a\ j} = \delta_{j, j_h} \ ,\\
\sum_a U_{ea} \beta^{(+)*}_{a\ j} = \delta_{j, j_l} \ ,
\end{eqnarray}
where $j_h$ is the index of the highest mass eigenstate (i.e. state 3 for the normal ordering and state 2 for the inverted ordering), and $j_l$ is the index of the lowest mass eigenstate (i.e. state 1 for the normal ordering and state 3 for the inverted ordering). Then, Eq. (\ref{phief}) becomes

\begin{eqnarray} \label{phiefs}
 \Phi^W_{e}(x)=\int \frac{d^3p}{(2\pi)^{3/2}}  & \Big[ \left( \chi^{(-)}(\vv{p}) a_{j_h}(\vv{p},-) + \sum_{a, j} U_{ea} \frac{m_a}{2P} \beta^{(+)}_{a\ j} \chi^{(+)}(\vv{p}) a_j(\vv{p},+) \right) e^{-i p \cdot x} \nonumber \\
 & + \left( \sum_{a, j} U_{ea} \frac{m_a}{2P} \alpha^{(-)*}_{a\ j} \chi^{(+)}(\vv{p})a^{\dagger}_j(\vv{p},-) -  \chi^{(-)}(\vv{p})a^{\dagger}_{j_l}(\vv{p},+) \right)e^{i p \cdot x}  \Big] \ .
\end{eqnarray}

One can then ask if these limits could change the propagator, Eq. (\ref{4propagator}), and consequently the decay half-life, Eq. (\ref{lifetime}).
It is preferable to calculate the contributions to electron neutrino propagator, Eq. (\ref{epropagator}), in an electron density medium using the full expression for the field $\Phi^M_e(x)$, Eq. (\ref{phief}), as

\small
\begin{equation}
\bra{0}  \Phi_{e}^W(x_1) \left( \Phi_{e}^{W}(x_2)\right)^T \ket{0} =
\int \frac{d^3p}{(2\pi)^3} \sum_{a,b,j} U_{ea} U_{eb} \Big[ \frac{m_b}{2P} \alpha^{(-)}_{a\ j} \alpha^{(-)*}_{b\ j}\chi^{(-)}(\vv{p})  \chi^{(+)T}(\vv{p}) - \frac{m_a}{2P} \beta^{(+)}_{a\ j} \beta^{(+)*}_{b\ j}\chi^{(+)}(\vv{p})  \chi^{(-)T}(\vv{p}) \Big] e^{-i p \cdot (x_1-x_2)} , 
\end{equation}
%\normalsize

\begin{equation}
-\bra{0}  \left( \Phi^{W}_{e}(x_2) \right)^T \Phi_{e}(x_1) \ket{0} =
\int \frac{d^3p}{(2\pi)^3} \sum_{a,b,j} U_{ea} U_{eb} \Big[ \frac{m_a}{2P} \beta^{(+)}_{a\ j} \beta^{(+)*}_{b\ j} \chi^{(-)}(\vv{p})  \chi^{(+)T}(\vv{p}) -   \frac{m_b}{2P} \alpha^{(-)}_{a\ j} \alpha^{(-)*}_{b\ j}\chi^{(+)}(\vv{p})  \chi^{(-)T}(\vv{p}) \Big] e^{i p \cdot (x_1-x_2)} .
\end{equation}
\normalsize
Using the unitarity of the $\alpha^{(-)}(P)$ and $\beta^{(+)}(P)$ matrices one gets

\begin{equation}
\bra{0}  \Phi_{e}^W(x_1) \left( \Phi_{e}^{W}(x_2)\right)^T \ket{0} =
 \sum_{a} U^2_{ea}   \int \frac{d^3p}{(2\pi)^3} \frac{m_a}{2P}  \big[ \chi^{(-)}(\vv{p})  \chi^{(+)T}(\vv{p}) - \chi^{(+)}(\vv{p})  \chi^{(-)T}(\vv{p}) \big] e^{-i p \cdot (x_1-x_2)} \ , 
\end{equation}

\begin{equation}
-\bra{0}  \left( \Phi^{W}_{e}(x_2) \right)^T \Phi_{e}(x_1) \ket{0} =
 \sum_{a} U^2_{ea}   \int \frac{d^3p}{(2\pi)^3} \frac{m_a}{2P}  \big[ \chi^{(-)}(\vv{p})  \chi^{(+)T}(\vv{p}) - \chi^{(+)}(\vv{p})  \chi^{(-)T}(\vv{p}) \big] e^{i p \cdot (x_1-x_2)} \ .
\end{equation}
Putting everything together on gets
\begin{equation} \label{new_propagator}
 \bra{0} T\left[ \Phi^W_{e}(x_1) \left( \Phi_{e}^W(x_2) \right)^T \right] \ket{0} = 
-i  \sum_a U^2_{ea} \int \frac{d^4p}{(2\pi)^4} \frac{m_a e^{-i p \cdot (x_1-x_2)}}{p^2-m^2_a+i \epsilon} \left(i \sigma^2\right) \ ,
\end{equation}
\end{widetext}
and using Eqs. (\ref{plc}) - (\ref{2propagator}) we recover the vacuum electron neutrino propagator of Eq. (\ref{4propagator}). Therefore, Eq. (\ref{lifetime}) for the inverse half-life stands.

The key element in the proof leading to this result is the approximation given in Eq. (2.21) of Ref. \cite{92giunti1557}, which lead to the unitarity of the $\alpha^{(-)}$ and $\beta^{(+)}$ matrices separately. Also, the unitarity of  $\alpha^{(-)}$ and $\beta^{(+)}$ matrices can be used to simplify the Majorana propagator of Eq. (\ref{epropagator}). One needs to investigate if the regular propagator for Majorana neutrino fields needed in cases where right handed currents are responsible for the neutrinoless double beta decay \cite{17ho-xv} may change the results. A quick look indicates that that might not be the case, but a more detailed analysis is needed. In addition, one would like to investigate the process of Majoron decay, where two neutrino propagators contribute. The similar analysis done for supernovae \cite{92giunti1557} indicates that some of the associated observables are affected by matter effects.

\section{Conclusions and Outlook} \label{section-conclusions}

In this paper we proposed a new paradigm, by considering the neutrino mixing in the high electron density existing in the atomic nuclei.  We showed that the standard phenomenology of neutrino emission and detection is obeyed. In particular we showed that, although the neutrinos(antineutrinos) are ``born`` in the highest(lowest) mass eigenstate inside the nucleus, they exit the atoms with the vacuum probabilities for each mass eigenstate. In addition, although the neutrinos(antineutrinos) are absorbed in the highest(lowest) mass eigenstate inside the nuclei, their vacuum probabilities are recovered after the transition  through the atomic electron cloud. We provided a simple explanation of these effects that relies on the extreme non-adiabatic evolution of the mixing in the atomic electron cloud. All the numerical simulations in the two-state approximation support this conclusions. Further numerical studies for the three state case are needed to investigate the regime where both adiabatic and non-adiabatic evolution of the mixing co-exists.

We also investigated the possible effect of the high electron density existing in the atomic nuclei on the neutrinoless double beta decay half-life for the mass mechanism, and we found that the effective neutrino mass parameter is the same as if the decay would take place in the vacuum. 

These results look simple and natural, but the road to them is complex. 
Our analysis provides a novel understanding of the neutrino phenomenology. 
Other observables, such as the Majoron decay rates in nuclei and supernovae \cite{92giunti1557,94berez439} may be significantly changed. 
The effects on the neutrino physics parameters of other decay modes need also be further investigated. 
%Our analysis shows that if one considers the effects neutrino mixing due to the high electron density in the atomic nuclei, the standard neutrino physics phenomenology remains unchanged.

\section{ACKNOWLEDGMENTS}
Support from the U.S. Department of Energy Grant No. DE-SC0015376 is acknowledged.

\bibliographystyle{apsrev}%
\bibliography{bb,nep}%

\end{document}